# Spin Transfer Torque in the Semiconductor/Ferromagnetic Structure in the Presence of Rashba Effect

Javad Vahedi[*] and Sahar Ghasab Satoory

*Department of Physics, Islamic Azad University, Sari Branch, Sari, Iran*

[*]Email: javahedi@gmail.com

**Abstract**

Spin transfer torque in magnetic structures occurs when the transverse component of the spin current that flows from the nonmagnetic medium to ferromagnetic medium are absorbed by the interface. In this paper, considering the Rashba effect on semiconductor region, we have discussed the spin transfer torque in semiconductor/ferromagnetic structure and obtained the components of spin-current density for two models: (I)-single electron and (II)- the distribution of electrons. We have shown that whatever the difference between Fermi surfaces in semiconductor and Fermi spheres for the up and down spins in ferromagnetic increase, the transmission probability decreases. The obtained results for the values used in this article illustrate that Rashba effect increases the difference between a Fermi sphere in semiconductors and Fermi sphere for the up and down spins in ferromagnetic. The results also show that the Rashba effect, brings an additional contribution to the components of spin transfer torque, which is not exist in the absence of the Rashba interaction. Moreover, the Rashba term has also different effects on the transverse components of the spin torque transfer.

## 1-Introduction

The concept of "spin transfer torque" was first introduced in 1996 independently by Slonczewski [1] and Berger [2]. Spin transfer means that, whenever a current of polarized electrons enters the ferromagnetic, there is a transfer of angular momentum between the electrons and the magnetic moment of the ferromagnetic area. In other words, the spin torque transfer occurs when the flow of transfer of spin angular momentum is not constant. For example, a spin current, which is polarized by filtration of a thin magnetic layer, faces a second polarization by another thin magnetic layer, which its magnetic moment direction is not alike the first one. During this process the second magnetic layer will definitely absorbs part of the spin angular momentum carried by the electron.

In the paper [3], authors have investigated and reported spin-current density and spin transfer torque on the general structure of a non-magnetic/ferromagnetic. The results indicate that only the transverse components of the spin-current play role on the spin transfer torque, while the longitudinal component of spin current does not show any effect. Here, the longitudinal component of current means the current which its electrons spin direction is in the same direction of the magnetization of the ferromagnetic and the transverse components of the current include current of electrons that their spin is normal to the magnetization direction (See figure 1).

The spin-orbit coupling in semiconductor low-dimensional systems plays an important role in the development of spintronic, which exploits electrons spin in order to store information [4, 5]. The prediction theory of spin transfer torque creates with harnessing the spin-orbit coupling first reported in reference [7]. And then experimentally for the first time were observed and reported in reference [8]. In Ref. [9] Authors have shown the effect of Rashba interaction using the Boltzmann transport theory that a maximum efficiency of spin torque transfer can be achieved by optimally changing parameter of the Rashba interaction. Also, authors of article [10] have reported spin transfer torque due to the non-uniform Rashba interaction. In this paper, considering the effect of Rashba in the semiconductor, we have investigated spin-current density and spin transfer torque in semiconductor Rashba/ferromagnetic junction for two free single electron and distribution of electrons models.

The paper is organized as follows. In section 2, we summarize the model, Hamiltonian and formalism. More details are given in the supplementary material. In section 3, we present our results by numerical calculations. Finally, conclusion is given in section 4.

## 2- Theory and model

The model considered is the Rashba/ferromagnetic semiconductor junction, in which the magnetization in the ferromagnetic layer has been considered in $\hat{z}$ direction. The Hamiltonian is given as follows:

$$H_{Ferro} = \begin{pmatrix} \frac{\hbar^2}{2m}k^2 & 0 \\ 0 & \frac{\hbar^2}{2m}k^2 \end{pmatrix} + \Delta \sigma_z \quad (1)$$

here $2\Delta$ is the exchange split and Hamiltonian in the Rashba semiconductor can be described as follows:

$$H = -\frac{\hbar^2 \vec{\nabla}^2}{2m} + \alpha\left(-i\vec{\nabla} \times \vec{E}\right) \cdot \vec{\sigma} \quad (2)$$

where $\alpha$ is the effective mass and $\vec{\sigma} = (\sigma_x, \sigma_y, \sigma_z)$ is the Paoli vector matrix. Assuming that electrical field is as $\vec{E} = (0, 0, E_z)$, we will have:

$$H_R = \begin{pmatrix} E_0 - \frac{\hbar^2}{2m}(\frac{\partial^2}{\partial x^2} - \frac{\partial^2}{\partial y^2}) & <\alpha E_z>(\frac{\partial}{\partial x} - i\frac{\partial}{\partial y}) \\ -<\alpha E_z>(\frac{\partial}{\partial x} + i\frac{\partial}{\partial y}) & E_0 - \frac{\hbar^2}{2m}(\frac{\partial^2}{\partial x^2} - \frac{\partial^2}{\partial y^2}) \end{pmatrix} \quad (3)$$

$$= \begin{pmatrix} \frac{\hbar^2}{2m}k^2 & i\Delta_R \\ -i\Delta_R^* & \frac{\hbar^2}{2m}k^2 \end{pmatrix}$$

which $\Delta R = \langle \alpha E_z \rangle (k_x - i k_y)$. In the above equation $<\alpha E_z>$ is the amount of expectation on the smallest subband with positive $E_0$ energy and normally, its observed amount is order of $10^{-11} eV\,m$ [6, 11]. Eigenstates for moving on the surface with wave vector of $\vec{k} = (k_x, k_y)$ and $s = \pm 1$ is given as:

$$\phi_{\vec{k}s}(\vec{r}) = N_{\vec{k}s} e^{i\vec{k}\cdot\vec{r}} \begin{pmatrix} is\left(\frac{k_x}{k} - i\frac{k_y}{k}\right) \\ 1 \end{pmatrix} \quad (4)$$

where $N_{\vec{k}s}$ is the normalization factor. Regarding the above equation for eigenstates, it is obvious that split subband by Rashba does not have spin polarization and then the electron energy dispersion can read as:

$$E_{\vec{k}s} = E_0 + \frac{\hbar^2}{2m}[(k+s k_R)^2 - k_R^2] \quad (5)$$

in which $k = \sqrt{k_x^2 + k_y^2}$ and $k_R = <\alpha E_z> m / \hbar^2$ is Rashba dependent wave vectors. In the following by using a rotation matrix:

$$U(\theta,\varphi) = \begin{pmatrix} \cos\theta/2 & e^{-i\varphi}\sin\theta/2 \\ e^{i\varphi}\sin\theta/2 & -\cos\theta/2 \end{pmatrix} \quad (6)$$

with $\varphi = -\pi/2, \theta = \pi/2$ and applying it to the Eq.(3), we will have new following Hamiltonian in the Rashba semiconductor:

$$U^\dagger(\theta,\varphi)H_R U(\theta,\varphi)=\begin{pmatrix}\frac{p_x^2}{2m}+\langle\alpha E_z\rangle k_x & 0 \\ 0 & \frac{p_x^2}{2m}-\langle\alpha E_z\rangle k_x\end{pmatrix} \quad (7)$$

Comparing Eq.(7) and Eq.(1), the Rashba semiconductor can be considered same as a pseudo-ferromagnetic which its magnetization is along the $\hat{z}$ direction. Now, the wave vector in a Rashba semiconductor [12] can be recast as follows:

$$k_R^s(q) = -s k_R + \sqrt{k_R^2 + k_F^2 - q^2}\; ; \; (s=\pm 1) \quad (8)$$

And the wave vector in ferromagnetic layer can be defined [8]:

$$k_{x,Ferro}^\sigma(q) = \sqrt{(k_F^\sigma)^2 - q^2} \quad (9)$$

In the two above equations, $k_R^s$ is the Rashba spin wave vector which shows that up and down spins in semiconductor region are different as much as $k_R$. Also the wave vector in this region in the presence and absence of Rashba is defined as follows, $k_x^R(q)=\sqrt{k_R^2+k_F^2-q^2}$, $k_x(q)=\sqrt{k_F^2-q^2}$, respectively. $k_{x,Ferro}^\sigma$ and $k_F^\sigma$ are spin dependent wave vector and Fermi wave vector in ferromagnetic. In order to calculate the charge current density and spin current density on the interface of Ferromagnetic and Rashba semiconductor, it is needed to introduce velocity operator in two areas. Assuming the propagation direction in $\hat{x}$ axis, the velocity operator in this direction is defined as follows:

$$v_x = \frac{\partial H_R}{\partial p_x}$$

$$= \begin{pmatrix} \frac{p_x}{m}+\frac{\hbar}{m}k_R & 0 \\ 0 & \frac{p_x}{m}-\frac{\hbar}{m}k_R \end{pmatrix} \quad (10)$$

this equation indicates that up and down spins in semiconductor move at different velocities in the propagation direction $\hat{x}$.

Now, in order to obtain transmission probability and eventually density of transmitted currents, it is needed to define wave functions in ferromagnetic and Rashba semiconductor regions. Therefore, we consider the wave functions for the Ferromagnetic/Rashba semiconductor junction (magnetization in the Ferromagnetic area is considered in $\hat{z}$ direction) in three dimensions as follows:

$$S=+1 \begin{cases} \psi_{Rashba}^+(x)=\left[e^{ik_+x}+r_\uparrow e^{-ik_-x}\right]e^{i\vec{q}.\vec{R}}|\uparrow\rangle \\ \psi_{Ferr}^+(x)=t_\uparrow e^{ik_x^\uparrow x} e^{i\vec{q}.\vec{R}}|\uparrow\rangle \end{cases} \quad (11)$$

$$S=-1 \begin{cases} \psi_{Rashba}^-(x)=\left[e^{ik_-x}+r_\downarrow e^{-ik_+x}\right]e^{i\vec{q}.\vec{R}}|\downarrow\rangle \\ \psi_{Ferr}^-(x)=t_\downarrow e^{ik_x^\downarrow x} e^{i\vec{q}.\vec{R}}|\downarrow\rangle \end{cases} \quad (12)$$

In these equation $q(k_y,k_z)$ and $R(y,z)$ are the wave vector and position vector, respectively. Reflection and transmission coefficients based on the boundary conditions, that is to say, wave function continuity and its first derivative at the interface:

$$\psi_{Rashba}^s(x=0)=\psi_{Ferro}^\sigma(x=0) \quad (13)$$

$$(\frac{\partial}{\partial x}\pm ik_R)\psi_{Rashba}^s(x=0)=\frac{\partial}{\partial x}\psi_{Ferro}^\sigma(x=0) \quad (14)$$

are as follows:

$$t_\sigma(q)=\frac{2\sqrt{k_R^2+k_F^2-q^2}}{\sqrt{k_R^2+k_F^2-q^2}+k_{Ferro}^\sigma(q)}; \quad (15)$$

$$r_\sigma(q)=\begin{cases} \dfrac{\sqrt{k_R^2+k_F^2-q^2}-k_{Ferro}^\sigma(q)}{\sqrt{k_R^2+k_F^2-q^2}+k_{Ferro}^\sigma(q)} & \text{if } q^2\leq(k_{Ferro}^\sigma)^2 \\[2ex] \dfrac{\sqrt{k_R^2+k_F^2-q^2}-ik_{Ferro}^\sigma(q)}{\sqrt{k_R^2+k_F^2-q^2}+ik_{Ferro}^\sigma(q)} & \text{if } q^2>(k_{Ferro}^\sigma)^2 \end{cases} \quad (16)$$

And the possibility of reflection and transmission:

$$R_\sigma(q)=|r_\sigma(q)|^2 \quad (17)$$

$$T_\sigma = \frac{k_x^\sigma(q)}{k_x^R(q)} |t_\sigma(q)|^2 \quad (18)$$

In general, when a polarized spin current with $k_x$ wave vector in a semiconductor layer gets into Ferromagnetic layer there is a possibility that by the interface, it undergoes under spin filtering and thus in Ferromagnetic layer, up and down spins feel different wave vector ($k_x^\downarrow, k_x^\uparrow$ respectively) [3].

There are three effective major factors contributing to the torque. The first factor is the spin filtering. This occurs when reflection probabilities are spin dependent. The second factor is the spin rotation. The spin rotation occurs when the quantity of $r_\uparrow^* r_\downarrow$ is not real and positive. So, this quantity can be written as follows:

$$r_\uparrow^* r_\downarrow = |r_\uparrow^* r_\downarrow| e^{i\Delta\varphi} \quad (19)$$

This relationship indicates that during the reflection, $\Delta\varphi$ phase is added to azimuthally angle $\varphi$ that shows the direction of incident electrons. In other words, the spin direction of electrons changes at the interface during reflection. This is completely a quantum phenomenon and cannot imagine a classical counterpart for that. Finally, the third factor in the development of spin transfer torque is spin precession. In the Ferromagnetic layer, as $k_x^\uparrow \ne k_x^\downarrow$, the spatially-varying phase factors which appear in the transmitted transverse spin currents, is in a way that their outcome creates a spatial precession.

$$\Delta k = k_x^\downarrow - k_x^\uparrow \quad (20)$$

Due to the conservation of angular momentum, the spin transfer torque on the studied material can be achieved getting help of calculation of pure spin current flow from this object surfaces [13].

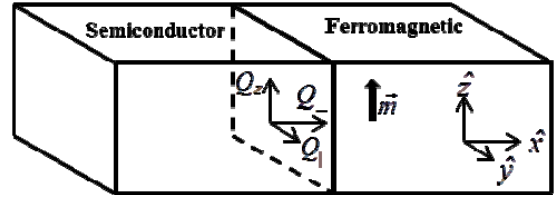

Figure-1 Sketch of the transverse and longitudinal components of spin-current density. In this figure, $Q_\perp$ and $Q_\parallel$ spin currents that are $Q_{xx}$ and $Q_{yx}$, respectively, are called the transverse components and $Q_z$ longitudinal spin current components.

$$N_{ST} = -\int_{pillbox\ surfaces} d^2 R\ \hat{n} \cdot \vec{Q}$$
$$= -\int_{pillbox\ volume} d^3 r\ \nabla \cdot \vec{Q} \quad (21)$$

However, it should be noted that a spin-current density $Q$ is a tensor, which includes spatial and spin parts. Point wise multiply of $\hat{n} \cdot \vec{Q}$ operates only on the spatial part and does not work on the spin part. This relationship is made of the fact that, if we consider a rectangular surface (Gauss surface) at the interface, the divergence theorem implies to apply a spin transfer torque of induced current to the interface magnetization.

In the following, above provided formulation will be investigated for the single-electron model and the distribution model of electrons.

## 2-1 Spin current for single electron

To investigate spin torque transfer in this model, first we define the spin current density as follows [3]:

$$Q(\mathrm{r}) = \sum_{i\sigma\sigma'} \mathrm{Re}\left[\psi_{i\sigma}^*(\mathrm{r}) \, s \otimes \hat{v} \, \psi_{i\sigma'}(\mathrm{r})\right]$$
$$= \frac{\hbar}{2} \sum_{i\sigma\sigma'} \mathrm{Re}\left[\psi_{i\sigma}^*(\mathrm{r}) \, \sigma \otimes \hat{v} \, \psi_{i\sigma'}(\mathrm{r})\right] \quad (22)$$

where $Q_{ij}(r)$ is a tensor quantity. Left index belongs to spin space and right index belongs to real space. The incident wave functions, reflection and transmission which are described by equations (11) and (12) can be written as a linear combination of spin up and spin down components $\psi = \psi_\uparrow + \psi_\downarrow$. Now, by inserting the wave functions in equation (22), current components will be as follows. For incident current component:

$$Q_\perp^{in} = Q_{xx}^{in} = \frac{\hbar^2 k_x}{m} \cos(2k_R x)$$
$$Q_\parallel^{in} = Q_{yx}^{in}(r) = \frac{\hbar^2 k_x}{m} \sin(2k_R x) \quad (23)$$
$$Q_{zx}^{in}(r) = \frac{\hbar^2 k_R}{m}$$

The reflected current:

$$Q_\perp^{ref} = Q_{xx}^{ref} = -\left[\frac{\hbar^2 k_x}{m}\right] \mathrm{Re}\left\{r_\uparrow^* r_\downarrow \, e^{i(2k_R)x}\right\}$$
$$= -\left[\frac{\hbar^2 k_x}{m}\right] |r_\uparrow^* r_\downarrow| \, \mathrm{Re}\left\{e^{i\Delta\varphi} e^{i(2k_R)x}\right\}$$
$$Q_\parallel^{ref} = Q_{yx}^{ref} = -\left[\frac{\hbar^2 k_x}{m}\right] \mathrm{Im}\left\{r_\uparrow^* r_\downarrow \, e^{i(2k_R)x}\right\} \quad (24)$$
$$= -\left[\frac{\hbar^2 k_x}{m}\right] |r_\uparrow^* r_\downarrow| \, \mathrm{Im}\left\{e^{i\Delta\varphi} e^{i(2k_R)x}\right\}$$
$$Q_{zx}^{ref} = -\frac{\hbar^2}{2m}\left\{[k_x + k_R]|r_\uparrow|^2 - [k_x - k_R]|r_\downarrow|^2\right\}$$

The transmitted current are

$$Q_\perp^{tr} = Q_{xx}^{tr} = \frac{\hbar^2}{2m} \mathrm{Re}\left[t_\uparrow^* t_\downarrow \, e^{i(k_x^\downarrow - k_x^\uparrow)x}\right](k_x^\uparrow + k_x^\downarrow)$$
$$Q_\parallel^{tr} = Q_{yx}^{tr} = \frac{\hbar^2}{2m} \mathrm{Im}\left[t_\uparrow^* t_\downarrow \, e^{i(k_x^\downarrow - k_x^\uparrow)x}\right](k_x^\uparrow + k_x^\downarrow) \quad (25)$$
$$Q_{zx}^{tr} = \frac{\hbar^2}{2m}\left[\left(k_x^\uparrow |t_\uparrow|^2 - k_x^\downarrow |t_\downarrow|^2\right)\right]$$

Now, substituting the obtained currents in Eq.(21), for the torque we have:

$$N_{st} = \left\{\frac{\hbar^2 k_x}{m} \cos(2k_R x) - \left[\frac{\hbar^2 k_x}{m}\right] \mathrm{Re}\left[r_\uparrow^* r_\downarrow e^{i(2k_R)x}\right]\right.$$
$$\left. - \frac{\hbar^2}{2m} \mathrm{Re}\left[t_\uparrow^* t_\downarrow \, e^{i(k_x^\downarrow - k_x^\uparrow)x}\right](k_x^\uparrow + k_x^\downarrow)\right\} \hat{x}$$
$$+ \left\{\frac{\hbar^2 k_x}{m} \sin(2k_R x) - \left[\frac{\hbar^2 k_x}{m}\right] \mathrm{Im}\left[r_\uparrow^* r_\downarrow e^{i(2k_R)x}\right]\right. \quad (26)$$
$$\left. - \frac{\hbar^2}{2m} \mathrm{Im}\left[t_\uparrow^* t_\downarrow \, e^{i(k_x^\downarrow - k_x^\uparrow)x}\right](k_x^\uparrow + k_x^\downarrow)\right\} \hat{y}$$
$$+ \left\{\frac{\hbar^2 k_R}{m} - \frac{\hbar^2}{2m} k_R (|r_\uparrow|^2 + |r_\downarrow|^2)\right\} \hat{z}$$

## 2.2 Spin current for distribution of electrons

Generally, in the description of transport, it is necessary to consider the effect of quantum mechanics coherence between all the electrons with different special modes. However, for the spin torque transfer model, up to now, experiments show [14, 15, 16] that it is not necessary to consider the coherence between electrons and different wave vectors of Fermi sources. This is equivalent to the use of semi-classical theory that only note the coherence between up and down spin state at any point of $\vec{k}$ in the Fermi sea.

In this model, the formula for calculating related to density of incident, reflected and transmitted currents are defined as follows [3]:

$$Q^{in} = \frac{\hbar}{2} \int_{v_x > 0} \frac{dk^3}{(2\pi)} \mathrm{Tr}\left[I^\dagger f(\vec{k}, \vec{r}) I \, \sigma\right] \otimes \vec{V}^{in}(\vec{k}) \quad (27)$$

$$Q^{ref} = \frac{\hbar}{2} \int_{v_x > 0} \frac{dk^3}{(2\pi)} \mathrm{Tr}\left[R^\dagger f(\vec{k}, \vec{r}) R \, \sigma\right] \otimes \vec{V}^{ref}(\vec{k}) \quad (28)$$

$$Q^{trans} = \frac{\hbar}{2} \int_{v_x > 0} \frac{dk^3}{(2\pi)} \mathrm{Tr}\left[T^\dagger f(\vec{k}, \vec{r}) T \, \sigma\right] \otimes \vec{V}^{trans}(\vec{k}) \quad (29)$$

which $f(\vec{k},\vec{r})$ is electron distribution function and would be defined as follows [3]:

$$f(\vec{k},\vec{r})=U(\vec{k},\vec{r})\begin{pmatrix} f_\uparrow(\vec{k},\vec{r}) & 0 \\ 0 & f_\downarrow(\vec{k},\vec{r}) \end{pmatrix} U^\dagger(\vec{k},\vec{r}) \quad (30)$$

In which $f_{\uparrow(\downarrow)}(\vec{k},\vec{r})$ is occupying function for up (down) spins. The operator $U(\vec{k},\vec{r})$ is the rotation matrix that has the following form:

$$U(\vec{k},\vec{r})=\begin{pmatrix} \cos(\theta/2)e^{-i\varphi/2} & -\sin(\theta/2)e^{-i\varphi/2} \\ \sin(\theta/2)e^{i\varphi/2} & \cos(\theta/2)e^{i\varphi/2} \end{pmatrix} \quad (31)$$

And for sake of simplicity, the dependence of $\vec{k}$ and $\vec{r}$ to angles $\theta$ and $\varphi$ have been removed. Now, with the help of Eqs.27, 28 and 29 the spin-current density components for $\theta=\pi/2$, $\varphi=-\pi/2$ at the interface between Rashba ferromagnetic and semiconductor, in the place of $\vec{r}=0$ are as following.

$$Q_{xx}^{in}=-\frac{1}{8\pi^2}\int dq\, q\, \Delta\mu\, \sin(2k_R x+\Delta\varphi) \quad (32)$$

$$Q_{xx}^{ref}=+\frac{1}{8\pi}\int \frac{dq\, q}{2\pi}\, |r_\uparrow^* r_\downarrow|\Delta\mu\, \sin(2k_R x+\Delta\varphi) \quad (33)$$

$$Q_{xx}^{trans}=-\frac{1}{4\pi}\int \frac{dq\, q}{2\pi}\, |t_\uparrow^* t_\downarrow|\Delta\mu\, \sin[(k_x^\downarrow-k_x^\uparrow)x]\left(\frac{v_x^\uparrow+v_x^\downarrow}{|2v_x|}\right) \quad (34)$$

$$Q_{yx}^{trans}=-\frac{1}{4\pi}\int \frac{dq\, q}{2\pi}\, |t_\uparrow^* t_\downarrow|\Delta\mu\, \cos[(k_x^\downarrow-k_x^\uparrow)x]\left(\frac{v_x^\uparrow+v_x^\downarrow}{|2v_x|}\right) \quad (35)$$

More details of calculations are given in the supplementary document.

## 3-Numerical results and discussion

In this section, we will investigate the spin wave vector, reflection and transmission coefficients for single-electron model, then the current density of the electron in distribution model will be studied.

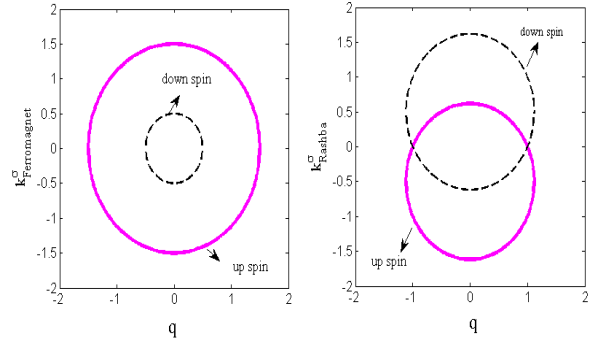

Figure 2: (Color online) $k_{Ferro}^\sigma$ spin wave vector for the up and down spins in ferromagnetic (left panel), $k_R^\sigma$ spin wave vector in Rashba semiconductor (right panel) versus $q$ for $k_F=1$, $k_x^\uparrow/k_F=1.5$ and $k_x^\downarrow/k_F=0.5$.

Figure-2 shows spin wave vector in the semiconductor and ferromagnetic layers. As it can be seen, the Rashba effect in semiconductors differs the wave vector for up and down spin, and this difference is such that the up and down spin wave vectors have the same amount, but each one to the non-Rashba are separated from each other as much as applied amount of Rashba. The difference in way of the spin wave vector detachment in Rashba semiconductor and ferromagnetic areas are clearly illustrated in Figure 2.

In Figure-3 we have displayed transmission and reflections probabilities for up and down spin in the absence and presence of Rashba interaction as a function of $q$. As it can be seen, possibility of reflection in the absence of Rashba for down spin in $q=0$ is close to zero and in $q\simeq 0.5$ reflection is one. While possibility of reflection for up spin in $q=0$ is almost zero and as the closer $q$ is to the Fermi wave vector ($k_F=1$), the complete reflection we will have.

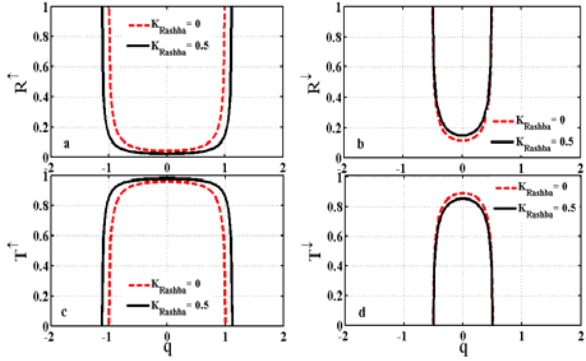

Figure-3 (Color online) Reflection and transmission possibility for up and down spins as function of $q$ for $k_F = 1$, $k_x^\uparrow / k_F = 1.5$ and $k_x^\downarrow / k_F = 0.5$.

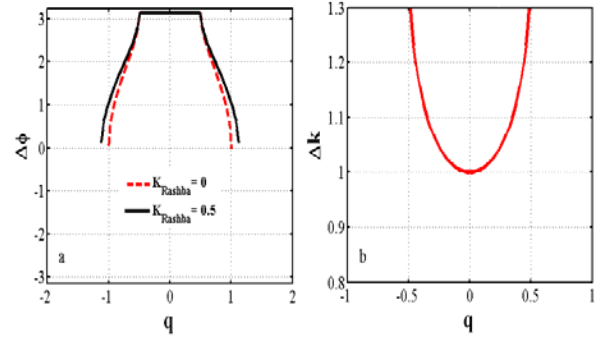

Figur-4 (Color online) Left panel shows the phase $\Delta\varphi$ acquired by an electron because its spin rotates upon reflection and right panel shows wave vector difference $\Delta k = k_x^\downarrow - k_x^\uparrow$ for a transmitted electron

Considering the Rashba interaction, the possibility reflection for up and down spin shows similar trend as the non-Rashba case. The reason is that wave vector increases in the semiconductor Rashba, considering the Rashba. In general, as the difference between the Rashba semiconductor wave vectors $(k_R^\sigma)$ and the ferromagnetic spin wave vector $(k_{Ferro}^\sigma)$ increases, the reflection increases as well.

Left panel in figure 4 presents phase difference of an electron due to spin rotation which experiences at the time of reflection. In the left panel of figure-4, at $q = 0$ we have $k_x^\uparrow = 1.5 k_F$ and $k_x^\downarrow = 0.5 k_F$. Thus, wave vector difference of $\Delta k$ is $k_F$ as it can be seen in the figure. Also, the wave vector difference of $\Delta k$ is 1.3, for $q = 0.5 k_F$. In fact, by reducing $q$ toward zero, the Fermi sphere difference for up and down spin increases, as a result $\Delta k$ reduces.

In the right panel of figure 4, we have illustrated wave vector difference of a transmitted electron. As it can be seen, by increasing $q$, phase difference in the presence and absence of Rashba reduces. In a way that in the absence of Rashba for $q = k_F = 1$, phase difference becomes zero, which means that the electrons around the Fermi surface with the same incident spin direction are reflected. While in the presence of Rashba, a non-zero phase difference in $q = k_F = 1$ is observed, even for small amounts of $k_R = 0.5$. In general, by increasing Rashba, phase difference of the reflected electrons compared to the incident electrons experience further changes around the Fermi surface.

In figure-5 we have plotted spin torque transfer in the presence and absence of Rashba interaction for a single electron model as function of distance from the interface of two environments. As one can see in the figure, in the absence of Rashba, the spin torque only contains components of transverse spin current. While considering the Rashba, torque, in addition to the components of the transverse, also will have longitudinal component. Moreover, the influence of Rashba interaction on the transverse components of the spin torque transfer is different.

Figure-6 shows density of spin-current for distribution of electrons as function of

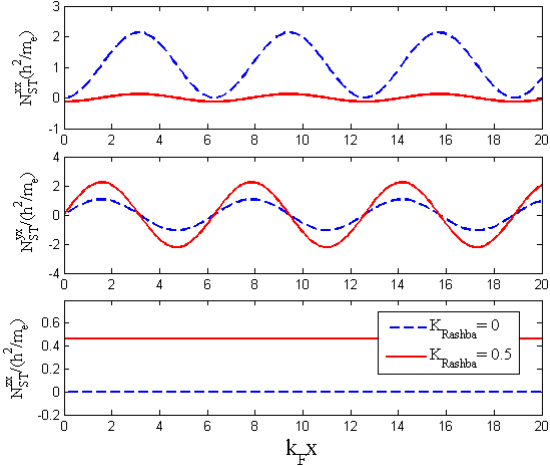

Figure-5: (Color online)Spin torque in the absence and presence of Rashba for a free single electron model. $k_F^\downarrow / k_F = 0.5$, $k_F^\uparrow / k_F = 1.5$

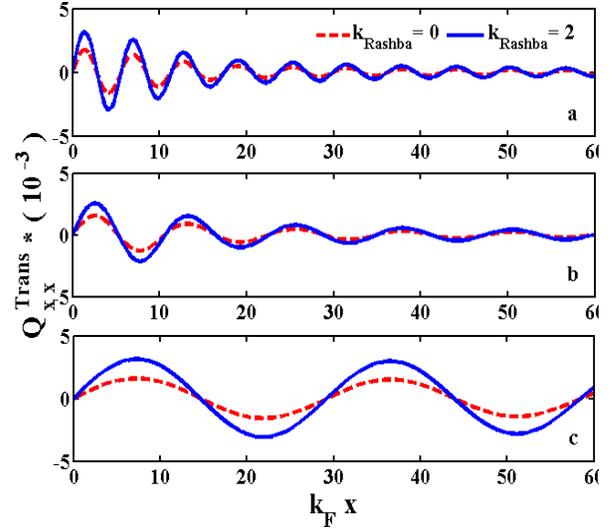

Figure-6: (Color online)Transmitted spin current as a function of distance from the interface for the Fermi energy of $k_F = 1$. a) $k_F^\downarrow / k_F = 0.5$, $k_F^\uparrow / k_F = 1.5$, b) $k_F^\downarrow / k_F = 0.5$, $k_F^\uparrow / k_F = 1.0$, c) $k_F^\downarrow / k_F = 0.9$, $k_F^\uparrow / k_F = 1.1$

distance from the interface of two environments for some parameters. Clearly, decay of the current density can be seen by getting distance from the interface. However, this decay is slightly small in the presence of Rashba interaction. In the absence of Rashba, Fermi wave vector for up and down spins are coincided. Thus, Fermi sphere in semiconductor is placed between Fermi spheres of up and down spins in ferromagnetic. In figure6-b, Fermi sphere of semiconductor is equal to Fermi sphere of spin majority in ferromagnetic and both are larger than Fermi sphere of spin minority. In figure 6-c Fermi spheres of majority and minority are slightly larger and smaller than Fermi sphere in semiconductor. Considering the Rashba effect leads the Fermi sphere and wave vector difference in a semiconductor $(k_x^R)$ with wave vector spin (majority and minority) in ferromagnetic $(k_x^\sigma)$ to be increased. The difference is less for up spin than down spin, as a result the chance of passing for that will increase. Thus, it can be concluded that in the transmitted current, in this structure, up spins play a major role in the transmitting current.

### 4-Conclusion

In this article we have discussed the theory of the spin transfer torque in the Rashba semiconductor/Ferromagnetic junction and obtained the components of spin-current density for two models: (I)- single electron model and (II)- the distribution of electrons. In the single electron model, the spin torque calculation, Eq.(26), shows that considering the Rashba effect in semiconductor and being different the Fermi wave vector for up and down spins in ferromagnetic region we will have all three components of torque. While in the absence of Rashba and by taking $k_x^\uparrow = k_x^\downarrow = k_x$, the spin torque will be only included the transverse components. In the distribution model of electrons, it can be concluded from the results of current density that the more difference between the Fermi sphere in semiconductor with Fermi spheres of spins majority and minority in ferromagnetic, the more transmitted current oscillation will get. In addition, considering the Rashba effect it leads to an increase in the transmission coefficient for up spins to down spin, while its current has similar trend to the current in the absence of the Rashba.